\documentclass[12pt,preprint]{emulateapj}
\usepackage{natbib,graphicx,epstopdf}

\shorttitle{\emph{Spitzer} Secondary Eclipse Photometry of WASP-4b}
\shortauthors{Beerer et al.}
\def\simle{\,\hbox{\hbox{$ < $}\kern -0.8em \lower 1.0ex\hbox{$\sim$}}\,}

\begin{document}

\title{Secondary Eclipse Photometry of WASP-4b with \emph{Warm Spitzer}} 

\author{ Ingrid M. Beerer \altaffilmark{1}, Heather A. Knutson \altaffilmark{1,2}, Adam Burrows \altaffilmark{3}, Jonathan J. Fortney \altaffilmark{4}, Eric Agol \altaffilmark{5}, David Charbonneau \altaffilmark{6},  Nicolas B. Cowan \altaffilmark{5}, Drake Deming \altaffilmark{7}, Jean-Michel Desert \altaffilmark{6}, Jonathan Langton \altaffilmark{8}, Gregory Laughlin \altaffilmark{4},  Nikole K. Lewis \altaffilmark{9},  Adam P. Showman \altaffilmark{9}  }

\altaffiltext{1}{Department of Astronomy, University of California at Berkeley, Berkeley, CA 94720}
\altaffiltext{2}{Miller Research Fellow}
\altaffiltext{3}{Department of Astrophysical Sciences, Princeton University, Princeton, NJ 05844}
\altaffiltext{4}{Department of Astronomy and Astrophysics, University of California at Santa Cruz, Santa Cruz, CA 95064}
\altaffiltext{5}{Department of Astronomy, University of Washington, Box 351580, Seattle, WA 98195}
\altaffiltext{6}{Harvard-Smithsonian Center for Astrophysics, Cambridge, MA 02138}
\altaffiltext{7}{Planetary Systems Laboratory, NASA's Goddard Space Flight Center, Greenbelt MD 20771}
\altaffiltext{8}{Department of Physics, Principia College, Elsah, IL 62028}
\altaffiltext{9}{Lunar and Planetary Laboratory, University of Arizona, Tucson, AZ 85721}

\begin{abstract}

We present photometry of the giant extrasolar planet WASP-4b at 3.6 and 4.5 $\mu$m taken with the Infrared Array Camera on board the \emph{Spitzer Space Telescope} as part of \emph{Spitzer's} extended warm mission. We find secondary eclipse depths of 0.319$\% \pm$0.031$\%$ and 0.343$\% \pm$0.027$\%$ for the 3.6 and 4.5 $\mu$m bands, respectively and show model emission spectra and pressure-temperature profiles for the planetary atmosphere. These eclipse depths are well fit by model emission spectra with water and other molecules in absorption, similar to those used for TrES-3 and HD 189733b. Depending on our choice of model, these results indicate that this planet has either a weak dayside temperature inversion or no inversion at all. The absence of a strong thermal inversion on this highly irradiated planet is contrary to the idea that highly irradiated planets are expected to have inversions, perhaps due the presence of an unknown absorber in the upper atmosphere. This result might be explained by the modestly enhanced activity level of WASP-4b's G7V host star, which could increase the amount of UV flux received by the planet, therefore reducing the abundance of the unknown stratospheric absorber in the planetary atmosphere as suggested in Knutson et al. (2010). We also find no evidence for an offset in the timing of the secondary eclipse and place a 2$\sigma$ upper limit on $|e\cos\omega|$ of 0.0024, which constrains the range of tidal heating models that could explain this planet's inflated radius.

\end{abstract}


\section{Introduction}

Observations of the emergent spectra from transiting extrasolar planets with the \emph{Spitzer Space Telescope}  have enabled us to probe the atmospheres of a class of giant extrasolar planets known as ``hot Jupiters''. These planets have masses and radii similar to the gas giants in our solar system, but orbit very close to their parent stars, with equilibrium temperatures ranging from 1000-2500 K. By measuring the wavelength-dependent decrease in light when the planet moves behind the star in an event known as a secondary eclipse, we can construct a dayside emission spectrum for the planet \citep{deming05, char05}. During its cryogenic mission, \emph{Spitzer} obtained multi-wavelength observations for fifteen extrasolar planets during secondary eclipse. The results of these studies indicate that hot Jupiter atmospheres can be distinguished by the presence or absence of a strong temperature inversion in the upper atmosphere \citep[e.g.,][]{burrows07b, burrows08, fortney08, barman08, madhu09}.

The \emph{Spitzer Space Telescope} is continuing to survey hot Jupiter emission spectra during its post-cryogenic mission. After its cryogen was exhausted in May 2009, only the 3.6 and 4.5 $\mu$m channels of the Infrared Array Camera (IRAC; Fazio et al. 2004) instrument are operational. Fortunately, these two wavebands are well placed to constrain the range of possible models for these atmospheres. Planets without a strong  inversion, which include HD 189733b \citep[e.g.][]{deming06, grillmair07, grillmair08, char08, barman08, swain09},  TrES-1 \citep{char05} and TrES-3 \citep{fressin10}, are best described by models that exhibit H$_2$O and CO absorption features, which cause a decrease in the eclipse depth at 4.5 $\mu$m relative to 3.6 $\micron$. A strong thermal inversion changes these features from absorption to emission, therefore increasing the flux at wavelengths greater than 4 $\micron$ in the atmospheres of planets, such as HD 209458b \citep{deming05, richardson07, knutson08, swain09}. Of the systems already observed with \emph{Spitzer}, eleven have been found to possess strong temperature inversions (see Knutson et al. 2010 for a review).

\begin{figure}\epsscale{1.0}
\plotone{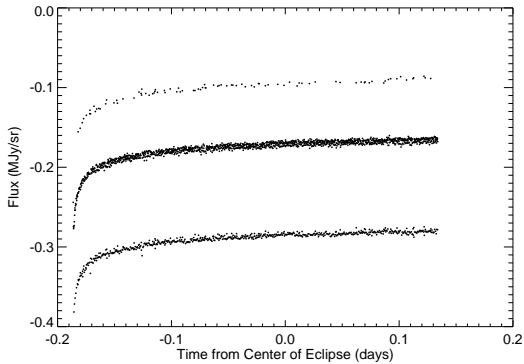}
\caption{Background estimate \emph{vs.} time for 3.6 $\mu$m images. We estimate the background by fitting a Gaussian to the central region of the histogram of counts in the entire array. The background estimates exhibit a ramp-like behavior, while also varying between three distinct levels.}
\end{figure}

In this paper, we present measurements of the transiting extrasolar planet WASP-4b spanning two times of secondary eclipse. WASP-4b is a 1.24 M$_{Jup}$ planet orbiting at 0.023 AU from a G7V star \citep{wilson08, gillon09, winn09}. If we assume that the planet absorbs all incident flux and re-emits that flux as a blackbody from the dayside alone, we calculate a maximum dayside effective temperature of about 2000 K. This highly irradiated planet provides an excellent test case for the correlation between temperature inversions and stellar irradiation (for a recent review see Wheatley et al. 2010). It has been hypothesized that absorbers such as gas-phase TiO in the upper atmosphere trap stellar irradiation, creating a thermal inversion \citep{hubeny03}. However, both because TiO is a heavy molecule and because titanium can condense into solid grains in night-side and day-side cold traps, significant macroscopic mixing would be required to maintain it in the upper atmosphere; it is not clear whether such vigorous mixing should be expected in a stably stratified atmosphere \citep{spiegel09}. This theory also fails to explain the presence of a temperature inversion in XO-1b's atmosphere, as this planet has a dayside temperature well below the condensation point for TiO. One alternative theory suggests that temperature inversions could be explained by absorption of UV and violet visible light by sulfur-containing species \citep{zahnle09}.

WASP-4b has a radius of 1.365$\pm$0.021 $R_{Jup}$ \citep{winn09}, which is larger than predicted by models of irradiated planets \citep{burrows07a, fortney07, guillot08}, placing it among a subset of ``bloated planets''. One possible explanation is that the inflated radius is caused by tidal heating due to ongoing orbital circularization. Using formulae from Liu et al. (2008), Winn et al. (2009) find that an orbital eccentricity between 0.002 and 0.02  would produce enough heat to inflate the planet to its observed size. Using radial velocity measurements, Mahusudhan $\&$ Winn (2009) find a 95.4$\%$ confidence upper limit on $e$ of 0.096. By measuring the time of secondary eclipse, we can place a much tighter upper limit on the parameter $e\cos\omega$, which will help determine whether tidal heating is a viable explanation.

In Sec. 2 we describe the observations and outline our fits to the data. In Sec. 3, we compare our results to the predictions of atmospheric models. Finally, in Sec. 4, we present our conclusions.

\begin{figure}\epsscale{1.0}
\plotone{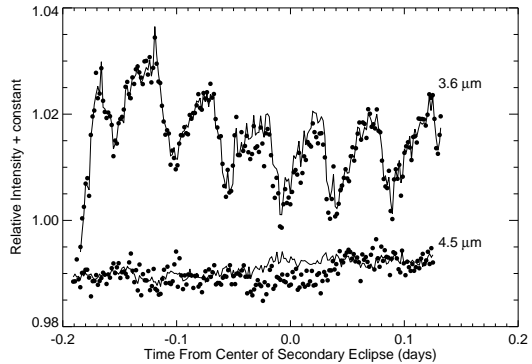}
\caption{Photometry at 3.6 and 4.5 $\mu$m \emph{vs.} time from center of secondary eclipse. The decorrelation functions to correct for intrapixel sensitivity are overplotted. We use a linear function of $x$ and $y$ position to correct the 4.5 $\mu$m photometry and a linear function of $x$, $y$ and time for the 3.6 $\mu$m observations.}
\end{figure}

\section{Observations and Methods}

We observed a secondary eclipse of WASP-4b in the 4.5 $\mu$m band on UT 2009 December 6 using IRAC on board the \emph{Spitzer Space Telescope}. We observed in full array mode with a 10.4 s integration time, yielding a total of 2115 images over a period of 7.7 hr. We observed a second secondary eclipse in the 3.6 $\mu$m band on UT 2009 December 9 using the same 10.4 s integration time, acquiring 2115 images in 7.7 hr.

We perform photometry on the Basic Calibrated Data (BCD) files produced by version S18.13.0 of the Spitzer pipeline. These data files are dark-subtracted, linearized, flat fielded and flux-calibrated. The cBCD images have been further corrected for artifacts due to bright sources, such as column pulldown, but these corrections have an unknown effect on time series photometry and we therefore elect to use the standard BCD images in our analysis. We extract the UTC-based Julian date for each image from the FITS header (keyword DATE$\_$OBS) and correct to mid-exposure. We convert to UTC-based BJD using the JPL Horizons ephemeris to estimate \emph{Spitzer}'s position during the observations.

We correct for transient ``hot pixels'' in a 20$\times$20 pixel box around the star by comparing each pixel's intensity to the median of the 10 preceding and 10 following frames at that position. If a pixel in an individual frame has an intensity $>$ 3$\sigma$ from the median value, its value is replaced by the median. We corrected 0.32$\%$ and 0.35$\%$ of the pixels in the box in the 3.6 $\mu$m and the 4.5 $\mu$m band images, respectively.

We estimate the background by fitting a Gaussian to the central region of the histogram of counts in the entire array. We find that the background varies significantly from frame to frame for both channels. The background values, which are plotted in Figure 1 for the 3.6 $\mu$m band images, display a ramp-like behavior, while also varying between three distinct levels. We find a similar pattern in channel 2. This behavior is likely a ubiquitous feature of warm \emph{Spitzer }, as it is also observed in the warm \emph{Spitzer} analysis of CoRoT-1 and CoRoT-2 \citep{deming10}.

We use three methods to measure the position of the star on the array. We calculate the flux-weighted centroid within 5.0 pixels of the approximate center of the star, fit a 2D Gaussian with a fixed width to a 7$\times$7 pixel subarray centered on the brightest pixel of the star \citep[e.g.,][]{agol10, stevenson10} and fit Gaussians to the marginal $x$ and $y$ sums using GCNTRD, which is part of the standard IDL astronomy library. Each method yields eclipse depths consistent to within 1$\sigma$. We found that using GCNTRD estimates for channel 1 and the 2D Gaussian estimates for channel 2 produced the smallest reduced chi-squared for the fits and therefore elect to use these position estimate methods. (2D Gaussian fits produced $\chi^2$=2535 and $\chi^2$=2239 for channel 1 and 2, respectively, whereas GCNTRD fits produced $\chi^2$=1973 and $\chi^2$=2273 for channel 1 and 2, respectively.) The difference in $\chi^2$ produced for the different position estimates is very large in channel 1. While we trim approximately the same number of points using both methods, we find that the GCNTRD positions result in a lower level of correlated noise in the final light curve. The rms difference between the 2D Gaussian and GCNTRD positions are 0.075 pixels in $x$ and 0.229 pixels in $y$ for channel 1 and 0.059 pixels in $x$ and 0.145 pixels in $y$ for channel 2. These differences are primarily in the form of a constant offset; we find that the relative change in position calculated using both methods is quite similar.

We perform aperture photometry with DAOPHOT using apertures ranging from 3.0 to 5.0 pixels in half pixel intervals. We carried out our fits for each of these apertures and found that the eclipse depths and times remain consistent for apertures between 3.0 and 5.0 pixels. We choose an aperture size of 3.5 for our analysis because it minimizes both the probability of hot pixels falling within the aperture and the root mean square (rms) scatter in the data. 

\begin{figure}\epsscale{1.0}
\plotone{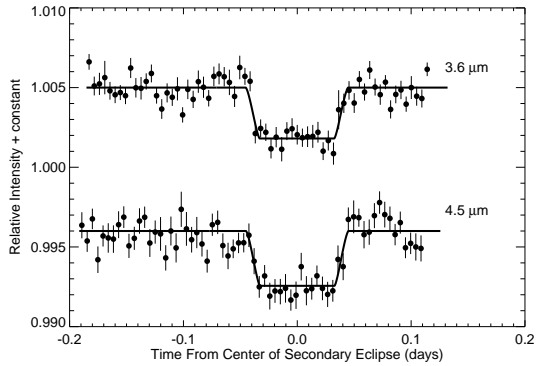}
\caption{Photometry at both wavebands after decorrelation \emph{vs.} time from center of secondary eclipse. The data are binned with 6.6 minute intervals. The error bars are based on the scatter of the individual points in each bin. The best-fit eclipse curve is overplotted.}
\end{figure}

\begin{deluxetable*}{ccccccrrrrr}
\tablecaption{Summary of Secondary Eclipse Results \label{summary}}
\tablewidth{0pt}
\tablenum{1}
\label{Secondary Eclipse Results}

\tablehead{
\colhead{Wavelength ($\mu$m)} & \colhead{Center of Eclipse (BJD)}  & \colhead{Depth ($\%$)} & \colhead{Eclipse Offset (min)} & \colhead{T$_{bright}$(K) \tablenotemark{a}} } 

\startdata
3.6 & 2455174.87731 $\pm$ 0.00087 & 0.319 $\pm$  0.031 & 0.5 $\pm$ 1.3 & 1832 $\pm$ 71\\
4.5 &  2455172.2011 $\pm$0.0013  & 0.343 $\pm$ 0.027 & 0.1 $\pm$ 1.9  & 1632 $\pm$ 56\\
\enddata
\tablenotetext{a}{We calculate the brightness temperature of the planet by finding the flux-weighted average of the planet-star flux ratio over each \emph{Spitzer} bandpass. We use a 5500 K PHOENIX NextGen model \citep{haus99} for the stellar spectrum and set the planet's emission spectrum equal to a blackbody, then solve for the temperature at which the planet-star flux ratio equals the observed eclipse depth.}

\end{deluxetable*}

The position of the star varies by 0.50 pixels in $x$ and 0.46 pixels in $y$ in the 3.6 $\mu$m images and by 0.21 pixels in $x$ and 0.26 pixels in $y$ in the 4.5 $\mu$m images. We discard any images where the measured flux, $x$ position or $y$ position was $>$ 3$\sigma$ from the median of the twenty frames surrounding the image in the time series. We removed a total of 10 images (0.47$\%$) and 15 images (0.71$\%$) from the 3.6 $\mu$m and 4.5 $\mu$m observations, respectively. 

The measured flux from the star varies significantly with its position on the pixel \citep[e.g.][]{char05,char08}. In order to correct for this intrapixel sensitivity, we fit the data with linear functions of the $x$ and $y$ positions. We fit the 4.5 $\mu$m data with a linear function of the form,

\begin{equation}
f = f_0(c_1(x-x_0)+c_2(y-y_0)+c_3)
\end{equation}

\noindent where $f$ is the flux measured on the array, $f_0$ is the original flux of the star, $x$ and $y$ are the positions of the star on the array, $x_0$ and $y_0$ are the median values of $x$ and $y$ over the time series and the constants $c_1$ - $c_3$ are free parameters. As a check we also try fits to the 4.5 $\micron$ data using a linear function of time instead of the $x$ and $y$ variables described above, but this results in noticeably poorer fits ($\chi^2$=2429 for the linear fit with four d.o.f. (degrees of freedom) and 2239 for the function of $x$ and $y$ including five d.o.f.). We also try a linear function of $x$, $y$ and time, but find the additional time term produces only a negligible improvement in the fit ($\chi^2$=2235, six d.o.f.). We fit the 3.6 $\mu$m data with a linear function in $x$, $y$, and time. We find that the linear fit in time produces a clear improvement in both the chi-squared value  ($\chi^2$=1973, six d.o.f. and $\chi^2$=2007, five d.o.f. for the fits with and without a linear fit for time, respectively) and the amount of correlated noise. We also try adding quadratic terms in $x$ and $y$ which are usually required when the star falls on the peak of the intrapixel curve (center of the pixel). However, we find that adding additional degrees of freedom in $x$ and $y$ has a negligible effect on the final time series, eclipse values, and chi-squared ($\chi^2$=1968, eight d.o.f.) and therefore elect to use the linear fit. Figure 2 shows the photometry with the decorrelation functions overplotted for each waveband.

We use a Markov Chain Monte Carlo method \citep{ford05, winn07} with $10^6$ steps to simultaneously determine the transit depth, timing of the eclipse, and the corrections for intrapixel sensitivity. We use five free parameters in the 4.5 $\mu$m data and six free parameters in the 3.6 $\mu$m data, including the linear term in time. We set the system parameters (planetary and stellar radii, orbital period, and orbital inclination) to the values given in Winn et al. (2009). We calculate the eclipse curve using the equations from Mandel $\&$ Agol (2002). The uncertainty for each point is set equal to the rms deviation of the out-of-eclipse data after removing the intrapixel effect. We also trim the first half hour of data from both the 3.6 and 4.5 $\mu$m time series because it exhibits larger deviations in position, perhaps due to settling of the telescope at a new pointing.

We take the median value of the distribution for each parameter as our best-fit solution. We calculate symmetric error bars about the median by finding the range over which the probability distribution contains 68$\%$ of the points above and below the median. The distributions for all parameters are nearly Gaussian and there are no strong correlations between parameters. Best-fit eclipse depths and times are shown in Table 1. As a check, we ran a second independent Markov chain for each channel and obtained identical results. Figure 3 shows the photometry after it has been corrected with the best-fit intrapixel correlation function with the best-fit eclipse curves overplotted.

We also calculate error bars using the `prayer-bead' method \citep{gillon09}. We divide our time series by the best-fit solution from the Markov chain, shift the time series in one point increments, multiply the best-fit solution back in and calculate the eclipse depth and time for the new data set. The prayer-bead distributions gave error bars that were consistent with the Markov chain errors and we elect to use the larger of the two errors in each case. For channel 1, we use the prayer-bead error for the eclipse depth (0.031\%) instead of the Markov error (0.019\%), whereas we use the Markov error for the time (1.3 min) instead of the prayer-bead error (0.72 min). For channel 2, we used the prayer-bead errors for both the eclipse depth (0.027\%) and time (1.9 min). The corresponding Markov errors are 0.023\% and 1.4 min.

We find that the rms variation in our light curve after correcting for intrapixel sensitivity is 1.1 and 1.2 times the predicted photon noise from the star at 3.6 and 4.5 $\mu$m, respectively. The reduced chi squared for our fits are 1.01 ($\chi^2$=1973, 1945 points, six d.o.f.) and 1.16 ($\chi^2$= 2239, 1941 points, five d.o.f.) for the 3.6 and 4.5 $\mu$m light curves, respectively. The error used in the fits is based on the rms deviation of the out-of-eclipse light curve rather than the predicted photon noise. Since we use the rms error estimates, the reduced $\chi^2$ should theoretically be equal to 1.0 if the noise is purely gaussian. The fact that we find reduced $\chi^2$ values exceeding 1.0 reflects the correlated noise present in our light curves which we take into account with the prayer-bead analysis.

\section{Discussion}

\subsection{Orbital Eccentricity}

The timing of the secondary eclipse is very sensitive to the planet's orbital eccentricity. Assuming a circular orbit and accounting for 23.4 seconds that light takes to travel across the orbit \citep{loeb05}, we would expect to see the secondary eclipse occur at a phase of 0.5002. In the event that there is significant advection of energy to the planet's night side we would expect an additional delay due to an offset hot spot on the planet's day side causing a change in the shape of ingress and egress. We estimate the maximum value of this delay to be 41 seconds based on a model in which the longitudinal advection time is 60\% of the radiative time, corresponding to a hot region shifted 30 degrees east of the substellar point \citep{williams06, cowen10}. We can use the difference between the predicted and observed orbital phase of secondary eclipse, including the light travel time but neglecting the unknown delay from a nonuniform surface brightness, to constrain $e\cos\omega$, where $e$ is the orbital eccentricity and $\omega$ is the argument of pericenter \citep[e.g.,][]{char05}.

\begin{figure*}\epsscale{0.9}
\plotone{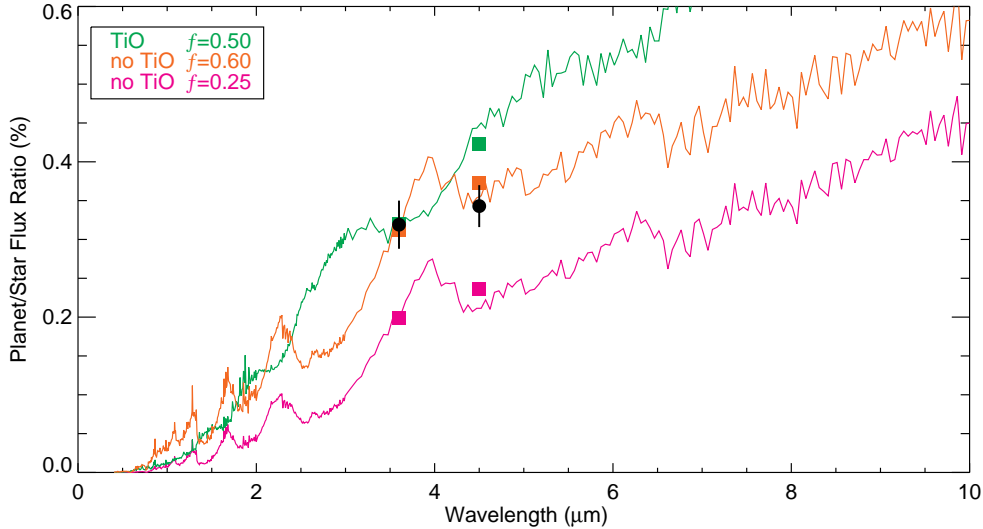}
\caption{Dayside planet/star flux ratio \emph{vs.} wavelength for three model atmospheres \citep{fortney08} with the band-averaged flux ratios for each model superposed (squares). The measured contrast ratios are overplotted (black circles). One model (\emph{green}) represents an atmosphere containing TiO in the upper atmosphere at equilibrium abundance. The other two models (\emph{orange} and \emph{magenta}) contain no TiO. The parameter $f$ represents the redistribution of energy over the planet's surface, where $f$=0.50 corresponds to dayside only redistribution and $f$=0.25 corresponds to uniform redistribution over the entire planet. We obtain the best fit to our measurements by a model with no TiO and little redistribution (\emph{orange}). }
\end{figure*}

\begin{figure}\epsscale{1.0}
\plotone{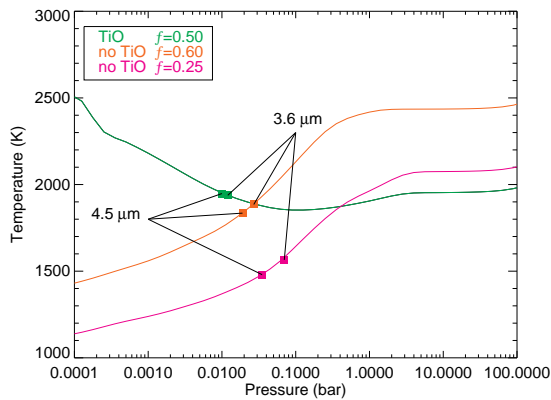}
\caption{ Dayside pressure-temperature profiles for the three model atmospheres in Figure 4 \citep{fortney08}. The green model contains TiO in the upper atmosphere and exhibits a strong temperature inversion for pressures below 0.01 bars. The orange and magenta profiles represent atmospheres with no TiO but have different values of the redistribution parameter $f$. The $f$=0.25 model has full redistribution of energy to the nightside, resulting in a cooler dayside profile. The hotter $f$=0.60 model provides the best fit to our measurements. We also indicate the approximate locations of the 3.6 and 4.5 $\micron$ photospheres (solid squares) for each model, estimated here as the pressure at which the model temperature matches the measured brightness temperature in each bandpass.}
\end{figure}

We find that the eclipse is offset from the predicted time based on the ephemeris from Winn et al. (2009) by 0.5$\pm$1.3 and 0.1$\pm$1.9 minutes in the 3.6 and 4.5 $\mu$m bands, respectively. We take the average of these two values weighted by the inverse of the variance and find a mean of 0.4$\pm$1.0 min, corresponding to $e\cos\omega = 0.00030 \pm 0.00086$. We place a 2$\sigma$ upper limit on $|e\cos\omega|$ of 0.0024, where we have calculated this limit by integrating over the histograms for the eclipse time. We integrate over the histograms from the Markov chain distribution for channel 1 and the prayer-bead distribution for channel 2. This upper limit implies that unless our line of sight happens to align very closely with the planet's major axis (i.e. the argument of pericenter $\omega$ is close to $\pi /2$ or $3\pi /2$) the orbit is nearly circular.

Ibgui, Burrows, \& Spiegel (2010) investigate the extra core power that would be needed to explain the otherwise anomalously large radius of WASP-4b. They find that approximately $7.8\times 10^{-8}$ L$_{\odot}$ of heating would be necessary for solar-metallicity opacity atmospheres, which decreases to $10^{-8}$ L$_{\odot}$ for 10$\times$ solar opacity atmospheres. Less power is necessary if the atmosphere helps retain more heat, as in the 10$\times$ solar case.  If this heating is due to tides, and the eccentricity is on the order of 0.001 and is maintained by an external planetary perturber in the system (as yet unidentified; Mardling 2007), then the Q$^{\prime}$ tidal dissipation parameter would be roughly between $3\times 10^4$ and $2\times 10^5$. In the Ibgui, Burrows, \& Spiegel (2010) paper, a value of 0.096 is assumed for the eccentricity and this leads them to derive a ``best-estimate" range for Q$^{\prime}$ between $3\times 10^8$ and $2\times 10^9$.  With our new constraint on the eccentricity of WASP-4b's orbit, and using the calculations of Ibgui, Burrows, \& Spiegel (2010), we now obtain a range for Q$^{\prime}$ that is more in line with the measured value of Jupiter of $10^5-10^6$ \citep{gold66,yoder81}.

\subsection{Atmospheric Temperature Structure}

In this paper we examine two distinct classes of hot Jupiter models. Figure 4 shows our planet/star contrast ratios and three models for the planetary atmosphere derived from one-dimensional, plane-parallel atmosphere codes following Fortney et al.$~$(2008). One model assumes the presence of the absorber TiO in the upper atmosphere at equilibrium abundances, whereas the two remaining model atmospheres contain no TiO. Fortney et al.$~$(2008) parameterize the unknown redistribution of energy to the planet's nightside by varying the stellar flux incident at the top of the planetary atmosphere by a geometric factor to account for dayside average ($f$=0.5) or planet-wide average ($f$=0.25) conditions. The slope between the 3.6 and 4.5 $\micron$ points on the model with TiO (\emph{green}) is much too steep to fit both measurements simultaneously. We find that WASP-4b is best fit by the orange model with no TiO (no inversion) and geometric factor $f$=0.60, resulting in a very hot dayside. This is a reasonable choice, as the projected area of the substellar point ($f$=1) is maximized during secondary eclipse while contributions from the cooler regions near the day-night terminator are correspondingly reduced, giving an average value of $f$=2/3 at opposition \citep{burrows08}. The dayside pressure-temperature profiles for these three models are displayed in Figure 5.

\begin{figure*}\epsscale{0.9}
\plotone{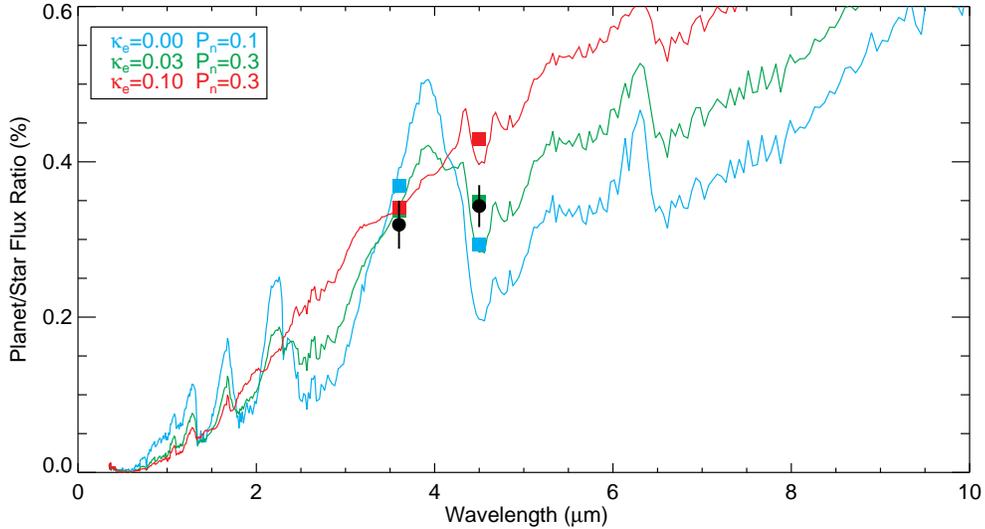}
\caption{Dayside planet/star flux ratio \emph{vs.} wavelength for three model atmospheres \citep{burrows08} with the band-averaged flux ratios for each model superposed (squares) to account for the widths of the \emph{Spitzer} bandpasses. The measured contrast ratios are overplotted (black circles). The blue model represents a non-inverted atmosphere ($\kappa_e$=0.0) with redistribution parameter $P_n$=0.1. An inverted atmosphere model is shown in red ($\kappa_e$=0.1), which exhibits water features in emission instead of absorption. The green model represents an atmosphere with a small amount of upper-atmosphere absorber, with optical opacity $\kappa_e$ equal to 0.03. The \emph{Spitzer} measurements are best matched by this model, which suggests that the atmosphere of WASP-4b has a moderate thermal inversion in its upper atmosphere.}
\end{figure*}

Figure 6 shows three models for the planetary atmosphere with greater degrees of freedom following Burrows et al. (2008). While the Fortney et al. models contain TiO in either zero or equilibrium abundance, the Burrows et al. models contain an unknown absorber at various optical opacities, parameterized by $\kappa_{e}$ in cm$^2/$g. Burrows et al. (2008) add a heat sink at a pressure range of 0.01 to 0.1 bars to model energy redistribution from the day to the nightside. As energy is most likely redistributed deep in the planetary atmosphere, this method for modeling heat transfer is physically motivated, but contains more degrees of freedom than the Fortney et al. models. General circulation models for these planets indicate that redistribution occurs continuously over a range of pressures \citep[e.g.,][]{showman08}, and we note that the range of pressures selected for our parameterized redistribution model can have a modest effect on the resulting pressure-temperature profiles, although it does not affect our main conclusions in this paper. The dimensionless parameter $P_n$ is a measure of the day to nightside energy redistribution, where $P_n$=0.0 represents no redistribution and $P_n$=0.5 represents full redistribution to the nightside.

Burrows et al. use a 5500 K Kurucz atmosphere model for the stellar spectrum \citep{kurucz79,kurucz94,kurucz05}, whereas Fortney et al. use a 5500 K PHOENIX NextGen model \citep{haus99}. As a check, we calculate the Burrows et al. planet-star flux ratio models using a PHOENIX NextGen stellar spectrum instead of the Kurucz spectrum and find that the differences are minimal and comparable to the differences caused by the uncertainty in the star's effective temperature. We find that the differences in the band-integrated flux-ratios using the two different stellar models vary between 0.003 and 0.005$\%$ and are therefore negligible when compared to our measurement errors.

We show an inverted atmosphere model (\emph{red}), with $\kappa_e$ and  $P_n$ set equal to the best-fit values for the archetype inverted atmosphere HD 209458b \citep{burrows07b, burrows08}. This inverted model is a poor fit to our measured contrast ratio at 4.5 $\mu$m. We find the best match is the model with a small amount of stratospheric absorber with $\kappa_e$=0.03 cm$^2/$g and relatively efficient day-night circulation with $P_n$=0.3. The band-integrated flux ratios for this model (\emph{green}) fall within 1$\sigma$ of the measured ratios in both bands. The pressure-temperature profiles in Figure 7 show that this best-fit model exhibits a modest temperature inversion for pressures below 0.01 bars, much weaker than the archetype inverted atmosphere HD 209458b. The blue non-inverted atmosphere model with parameters $\kappa_e$=0.0 and $P_n$=0.1 fails to fit our measurements at both wavebands.

We note that while the best-fit Burrows et al. model indicates moderately efficient ($P_n=0.3$) day-night circulation, the best-fit Fortney et al. model with $f=0.60$ requires minimal day-night circulation. It is perhaps not surprising that these relatively simple models disagree, given the differences in their treatment of the incident flux, optical opacities, and energy loss (if any) to the night side. We find some tentative evidence that this disagreement may be systematic, as published results for HD 189733b \citep{knutson09a} and HD 209458b \citep{fortney10} from Fortney et al. favor a hot dayside whereas Burrows et al. models predict greater energy redistribution ($P_n=0.30$ and 0.15 for HD 209485b and HD 189733b, respectively; Burrows et al. 2007b, 2008).  Multi-wavelength phase curve observations allow us to test these predictions, at least for the brightest systems (e.g., Knutson et al. 2007, 2009a).

\begin{figure}\epsscale{1.0}
\plotone{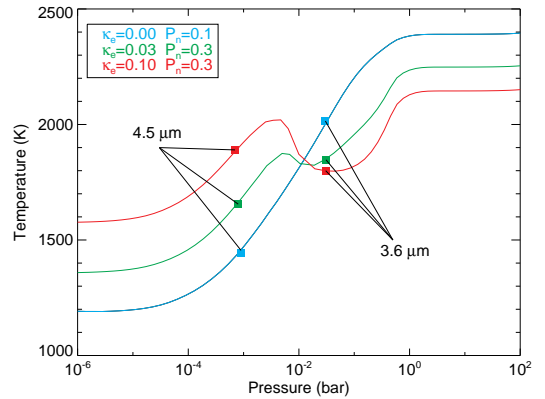}
\caption{ Dayside pressure-temperature profiles for three model atmospheres with various values of the parameters $P_n$ and $\kappa_{e}$ \citep{burrows08}. The blue model represents an atmosphere with no inversion. The red model corresponds to an atmosphere with an additional absorber with optical opacity $\kappa_e$=0.1 cm$^2/$g. The absorber, which is added high up in the atmosphere where the pressure is below 0.03 bars, traps stellar irradiation and creates a temperature inversion. The green model with $\kappa_e$=0.03 and $P_n$=0.3 provides the best fit to our measurements of WASP-4b. This model exhibits a slight temperature inversion for pressures less than 0.01 bars. Burrows et al. (2008) add a heat sink at a pressure range of 0.01 to 0.1 bars to model energy redistribution from the day to the nightside, which contributes to the decrease in dayside temperatures between 0.05 and 1.0 bars for the $P_n$=0.3 models. We also indicate the approximate locations of the 3.6 and 4.5 $\micron$ photospheres (solid squares) for each model, estimated here as the median pressure of the $\tau=$2/3 surface over the range of wavelengths spanned by each bandpass. We find the same approximate photosphere locations by solving for the pressure at which the the temperature of the model matches the measured brightness temperature in each band. Due to the width of the \emph{Spitzer} bandpasses, we actually see flux from a wide range of pressures. Typical ranges are 7$\times$10$^{-3}-2\times$10$^{-1}$ bars and 2$\times$10$^{-4}-1\times$10$^{-1}$ bars at 3.6 and 4.5 $\micron$, respectively.}
\end{figure}

 \emph{Spitzer} infrared observations indicate that the atmospheres of extrasolar giant planets tend to exhibit properties ranging between two differing types, exemplified by HD 189733b, whose emission spectrum features water and other molecules in absorption, and HD 209458b, which exhibits these features in emission. Table 2 shows the published values of $\kappa_e$ and $P_n$ for a range of planets with \emph{Spitzer} observations; WASP-4b is similar to HD 189733b and TrES-3 in that it requires a relatively small amount of absorber as compared to HD 209458b.

Assuming WASP-4b absorbs with zero albedo and re-emits on the dayside only, the planet's predicted dayside effective temperature is approximately 2000 K. If the planet emits uniformly over both hemispheres, we would expect an effective temperature of about 1650 K. We fit both measured eclipse depths simultaneously using a 5500 K PHOENIX NextGen model \citep{haus99} for the stellar spectrum and a blackbody for the planet's spectrum, and find that WASP-4b has a best-fit blackbody temperature of 1700 K. Given such high irradiation, it is somewhat surprising that WASP-4b exhibits at most a relatively weak thermal inversion. WASP-4b is therefore an exception to the general trend that highly irradiated planets are more likely to have strong thermal inversions.

In Knutson et al. (2010) we propose that there exists a correlation between temperature inversions and the activity levels of the host star, where the increased UV flux from active host stars destroy the compounds that are responsible for producing temperature inversions. We use \ion{Ca}{2} H \& K line strengths as indicators of stellar activity levels. In Knutson et al. (2010) we obtain Keck HIRES spectra for WASP-4b and find the \ion{Ca}{2} H \& K line strength estimates are $S_{HK}$=0.194 and $\log\left(R_{HK}\prime\right)$=$-$4.865, assuming a model $B-V$ color of 0.74 for a 5500 K star. These line strengths indicate that WASP-4 is a moderately active star, with a $\log\left(R_{HK}\prime\right)$ value that falls near the division between classes. However, WASP-4b's smaller orbital distance relative to HD 189733b means that it intercepts proportionally more of its star's flux, and as a result we estimate that the UV flux per unit area incident at the surface of WASP-4b is approximately half that received by the planet HD 189733b and twice that received by WASP-2 (see discussion in Knutson et al. 2010). 

We also calculate a value for the empirical index defined in Knutson et al. (2010) as the difference between the slope across the measured 3.6 and 4.5 $\micron$ eclipse depths and the slope of the best-fit blackbody function for the planet, which provides an observational means to distinguish between the two hot Jupiter atmosphere types.  We find a value of $-$0.09$\pm$0.04 in this index for WASP-4b, which suggests that this planet is best classified in the same type as HD 189733b (index of $-$0.15$\pm$0.02) and TrES-3b (index of $-$0.10$\pm$0.05).  Planets with strong inversions typically have positive values in this index, therefore this result is consistent with our earlier conclusion that WASP-4 displays at most a relatively weak temperature inversion.

\begin{deluxetable}{ccccccrrrrr}
\tablecaption{Effective Temperature and Burrows et al. Model Parameters for Extrasolar Giant Planets  }
\tablewidth{0pt}
\tablenum{2}
\label{Secondary Eclipse Results}

\tablehead{
\colhead{Name} & \colhead{$T_{eff}$ (K) \tablenotemark{a}}  & \colhead{$\kappa_e$} & \colhead{$P_n$} & \colhead{Reference} } 

\startdata
TrES-3      &  2000 & 0.01 & 0.3 & Fressin et al. 2010\\
WASP-4b     &  2000  & 0.03 & 0.3 & this paper\\
HD 189733b     &  1400  & 0.035 & 0.15 & Grillmair et al. 2008\\
HD 209458b    &  1700  & 0.1 & 0.3 & Burrows et al. 2007b\\
TrES-4     &  2100  & 0.1 & 0.3 & Knutson et al. 2009b \\
XO-1b     &  1400  & 0.1 & 0.3 & Machalek et al. 2008\\
XO-2b     &  1600  & 0.1 & 0.3 & Machalek et al. 2009\\
TrES-2    &  1800  & 0.3 & 0.3 & Spiegel \& Burrows 2010\\
HAT-P-7b     &  2500  & 1.1 & 0.0 & Spiegel \& Burrows 2010 \\
\enddata
\tablenotetext{a}{Predicted blackbody temperature for the planet assuming an albedo of zero and no nightside redistribution of energy.}

\end{deluxetable}

\section{Conclusions}

We observed two secondary eclipses of the extrasolar planet WASP-4b at 3.6 and 4.5 $\mu$m as part of \emph{Spitzer}'s extended warm mission. By measuring the time of the eclipse, we estimate a 2$\sigma$ upper limit on the parameter $|e\cos\omega|$ of 0.0024. This limit implies that unless our line of sight happens to align closely to the planet's major axis, the planet's orbit must be nearly circular. Although this upper limit does not rule out tidal heating, it constrains the range of tidal heating models that could explain this planet's inflated radius.

We find secondary eclipse depths of 0.319$\% \pm$0.031$\%$ and 0.343$\% \pm$0.027$\%$ for the 3.6 and 4.5 $\mu$m bands, respectively. These results are consistent with a spectrum exhibiting water and CO in absorption. We find that the atmosphere can be well characterized by models with a modest or no thermal inversion. Measurements at other wavelengths would help to distinguish between these two models. The absence of a strong thermal inversion makes WASP-4b an exception to the rule that inversions are found on planets that receive higher stellar irradiation. Other exceptions include the highly irradiated extrasolar planet TrES-3 \citep{fressin10} which does not have a temperature inversion and XO-1b \citep{mach08} which possesses a temperature inversion despite being relatively cool. These planets indicate that there must exist additional stellar or planetary parameters, other than equilibrium temperature, responsible for determining the relative strengths of thermal inversions in hot Jupiter atmospheres.

This work demonstrates that Warm \emph{Spitzer}, which operates with the 3.6 and 4.5 $\mu$m channels only, can be successfully used to characterize the properties of hot Jupiter atmospheres.  The increasing availability of ground-based eclipse detections in the near-IR \citep[e.g.,][]{gillon09, croll10a, croll10b, gibson10, lopez10} will also help to resolve ambiguities in the interpretation of the Spitzer data for many of these planets. Indeed, our models predict that WASP-4b should have an eclipse depth of 0.1$-$0.2\% in the $Ks$ band (2.15 microns). Croll et al. (2010b) measured a secondary eclipse depth of $0.133^{+0.018}_{-0.016}$ in this same bandpass for TrES-3, which has an apparent brightness and other properties similar to those of WASP-4. The TrES-3 K-band detection augers well for a similar WASP-4b measurement, which would provide a further point of comparison for the atmospheric models that we present. By the end of its post-cryogenic mission, \emph{Spitzer} will observe more than twenty systems during secondary eclipse. When combined with the nineteen systems observed during the cryogenic mission, as well as any available ground-based detections,  these results will allow us to search for correlations with other system parameters that could provide valuable clues to the origin of temperature inversions in hot Jupiter atmospheres.

\acknowledgments
This work is based on observations made with the \emph{Spitzer Space Telescope}, which is operated by the Jet Propulsion Laboratory, California Institute of Technology, under contract with NASA. Support for this work was provided by NASA. Heather A. Knutson is supported by a fellowship from the Miller Institute for Basic Research Science. Eric Agol acknowledges the support of NSF CAREER Grant No. 0645416.

\newpage

\end{document}